\documentclass[12pt,preprint]{aastex}

\usepackage[]{natbib}

\usepackage{longtable}

\usepackage{lscape}

\usepackage{fullpage}

\usepackage{epsfig}

\usepackage{graphicx}

\usepackage{epstopdf}

\usepackage{amssymb}

\usepackage{lineno}

\usepackage{color,soul}

\usepackage{graphics}

\begin{document}





\title{Detection of Rotational Spectral Variation on the M-type asteroid (16) Psyche}


\author{Juan A. Sanchez\altaffilmark{1}}
\affil{Planetary Science Institute, Tucson, AZ 85719, USA}
\email{jsanchez@psi.edu}

\author{Vishnu Reddy\altaffilmark{1}}
\affil{Lunar and Planetary Laboratory, University of Arizona, Tucson, AZ 85721, USA}

\author{Michael K. Shepard}
\affil{Bloomsburg University, Bloomsburg, PA 17815, USA}

\author{Cristina Thomas\altaffilmark{1}}
\affil{Planetary Science Institute, Tucson, AZ 85719, USA}

\author{Edward A. Cloutis}
\affil{Department of Geography, University of Winnipeg, Winnipeg, Manitoba, Canada}

\author{Driss Takir\altaffilmark{1}}
\affil{Astrogeology Science Center, U.S. Geological Survey, Flagstaff, AZ 86001, USA}

\author{Albert Conrad}
\affil{LBT Observatory, University of Arizona, Tucson, AZ 85721, USA}

\author{Cain Kiddell}
\affil{Department of Geography, University of Winnipeg, Winnipeg, Manitoba, Canada}

\author{Daniel Applin}
\affil{Department of Geography, University of Winnipeg, Winnipeg, Manitoba, Canada}

\altaffiltext{1}{Visiting Astronomer at the Infrared Telescope Facility, which is operated by the University of Hawaii under Cooperative Agreement no. NNX-08AE38A with 
the National Aeronautics and Space Administration, Science Mission Directorate, Planetary Astronomy Program.}

\begin{abstract}

The asteroid (16) Psyche is of scientific interest because it contains $\sim$ 1\% of the total mass of the asteroid belt and is thought to be the remnant metallic core of a protoplanet. Radar observations 
have indicated the significant presence of metal on the surface with a small percentage of silicates. Prior ground-based observations showed rotational variations in the near-infrared 
(NIR) spectra and radar albedo of this asteroid. However, no comprehensive study that combines multi-wavelength data has been conducted so far. Here we present rotationally resolved NIR spectra (0.7-2.5 $\mu$m) of (16) Psyche obtained with the NASA 
Infrared Telescope Facility. These data have been combined with shape models of the asteroid for each rotation phase. Spectral band parameters extracted from the NIR spectra show that 
the pyroxene band center varies from $\sim$ 0.92 to 0.94 $\mu$m. Band center values were used to calculate the 
pyroxene chemistry of the asteroid, whose average value was found to be Fs$_{30}$En$_{65}$Wo$_{5}$. Variations in the band depth were also observed, with values ranging from 1.0 to 1.5\%. Using a new laboratory spectral calibration we estimated an 
average orthopyroxene content of 6$\pm$1\%. The mass-deficit region of Psyche, which exhibits the highest radar albedo, also shows the highest value for spectral slope and the minimum band depth. The spectral characteristics of Psyche suggest that its parent body did not have the typical structure expected for a 
differentiated body or that the sequence of events that led to its current state was more complex than previously thought.

\end{abstract}

\keywords{minor planets, asteroids: general --- techniques: spectroscopic}


\section{Introduction}

The asteroid (16) Psyche is often regarded as the archetype of M-type asteroids, which were originally defined by \cite{1976AJ.....81..262Z} and later included in the Tholen taxonomic 
system \citep{1984PhDT.........3T}. They are characterized as having moderate albedos ($\sim$ 0.1-0.3) and featureless red-sloped spectra in the visible wavelength 
region (0.3-1.1 $\mu$m). Among these objects, Psyche is the largest, with a diameter of 226$\pm$15 km \citep{2017Icar..281..388S}. It has a rotation period of $\sim$ 4.2 h \citep{2002Icar..159..369K} 
and a visual albedo of 0.12 \citep{2002AJ....123.1056T}. Early observations showed similarities between Psyche and iron meteorites \citep{1973Icar...19..507C, 1979aste.book..655C}, which led to 
the assumption that this asteroid is the remnant metallic core of a differentiated body that was catastrophically disrupted \citep[e.g.,][]{1989aste.conf..921B}. Numerous 
observations of Psyche using different techniques seem to support this theory. Radar observations have shown that Psyche has a radar albedo almost three times as high as the average 
value measured for S- and C-type asteroids \citep{2007Icar..186..152M, 1985Sci...229..442O, 2008Icar..195..184S, 2017Icar..281..388S}, consistent with an object that is predominantly metallic. 
\citet{2002A&A...395L..17K} studied the perturbing effects caused by Psyche on the motion of 13206 Baer (1997 GC22) during a close encounter between these two asteroids. This yielded a 
density estimation of 6.98$\pm$0.58 g cm$^{-3}$ for Psyche, a value that closely resembles that of iron meteorites (7.5 g cm$^{-3}$). Similarly, estimations by \cite{2017Icar..281..388S} 
from radar observations led to a bulk density of 4.5$\pm$1.4 g cm$^{-3}$ for this asteroid, consistent with an Fe-Ni composition and 40$\%$ macroporosity. \cite{2013Icar..226..419M} estimated Psyche's thermal inertia from interferometric observations obtained with the VLT Interferometer, finding values of 133 $J$ $m^{-2}$ $s^{-0.5}$ $K^{-1}$ and 
114 $J$ $m^{-2}$ $s^{-0.5}$ $K^{-1}$ (no roughness and low roughness, respectively). These high values were considered as evidence of a metal-rich surface for Psyche 
\citep{2013Icar..226..419M}. In the NIR wavelength range ($\sim$ 0.7-2.5 $\mu$m) the spectrum of Psyche exhibits a red slope and a weak absorption band at 
$\sim$ 0.9 $\mu$m \citep{2004AJ....128.3070C, 2005Icar..175..141H, 2010Icar..210..674O}. These spectral characteristics are consistent with a surface metal component and a low-Fe 
pyroxene phase. \cite{1995Icar..117..443B} carried out rotationally resolved observations of Psyche in the visible wavelength 
range (0.5-0.9 $\mu$m), finding no significant variations within a precision of 1\%. However, \cite{2008Icar..195..206O} reported variations in the spectral slope 
and the 0.9 $\mu$m band depth from NIR spectra (0.8-2.5 $\mu$m) of Psyche.

The physical characteristics and origin of Psyche make this asteroid an interesting case for study, as demonstrated not only by the extensive work carried out so far, but 
also by the prospect of an unmanned mission to this object \citep{2014LPI....45.1253E, 2015LPI....46.1632E}. Here we present new rotationally resolved NIR spectra of 
(16) Psyche. In section 2 we describe the observations and data reduction procedure. In section 3 the NIR spectra are combined with shape models of the asteroid 
for each rotation phase. Band parameters are extracted from the spectra and used to determine the pyroxene chemistry and abundance across the surface of the 
asteroid. The effects of the phase angle, grain size, and space weathering on the spectra are also discussed. The global composition of Psyche and possible formation 
scenarios for this asteroid are discussed in section 4. Finally, in section 5 we summarize our main findings.

\section{Observations and data reduction}

NIR spectra ($\sim$ 0.7-2.5 $\mu$m) of (16) Psyche were obtained on three nights between December 2015 and February 2016 using the SpeX instrument \citep{2003PASP..115..362R} on 
the NASA Infrared Telescope Facility (IRTF) on Mauna Kea, HawaiÕi. The spectra were obtained using SpeX in its low-resolution (R$\sim$150) prism mode with a 0.8" slit width. A G-type local extinction star was observed 
before and after the asteroid in order to correct for telluric features. Spectra of the solar analog star Hyades 64 (SAO 93936) were also acquired to correct for spectral slope variations that could arise due to the use of a 
non-solar extinction star. During the observations, the slit was oriented along the parallactic angle in order to reduce the effects of differential atmospheric refraction.
The observational circumstances are presented in Table \ref{t:Table1}. Data reduction was performed using the IDL-based software 
Spextool \citep{2004PASP..116..362C}. A detailed description of the steps followed in the data reduction process can be found in \cite{2013Icar..225..131S, 2015ApJ...808...93S}.

\begin{table}[!ht]
\begin{center}\footnotesize
\caption{\label{t:Table1} {\footnotesize Observational circumstances. The columns in this table are: Date (UTC), Mid UTC, sub-Earth latitude, body-centered longitude, 
rotation phase, phase angle ($\alpha$), V-magnitude and airmass. Rotation phases are based on the shape model and are calculated by subtracting the 
longitude to 360$^\mathrm{o}$.}}

\rule{0pt}{2ex}

\hspace*{-1.9cm}
\begin{tabular}{crrrrrrrrrrr}
\tableline\tableline

Date (UTC)&Mid UTC&Lat. ($^\mathrm{o}$)&Long. ($^\mathrm{o}$)&Rot. Phase ($^\mathrm{o}$)&$\alpha$ ($^\mathrm{o})$&Mag. (V)&Airmass  \\

12/08/2015&10:30:38&-46.74&346&14&1.8&9.4&1.00 \\
12/08/2015&11:43:40&-46.75&242&118&1.8&9.4&1.06 \\
12/08/2015&12:47:12&-46.76&151&209&1.8&9.4&1.20 \\
12/09/2015&14:01:48&-46.99&145&215&1.7&9.4&1.63 \\
02/10/2016&05:10:23&-52.19&70&290&19.5&10.8&1.02 \\
02/10/2016&05:30:41&-52.19&41&319&19.5&10.8&1.01 \\
02/10/2016&06:34:30&-52.19&309&51&19.5&10.8&1.02 \\
02/10/2016&08:05:17&-52.18&180&180&19.5&10.8&1.17 \\

\tableline
\tableline
\end{tabular}\hspace*{-1.8cm}
\end{center}
\end{table}

\section{Results}

\subsection{Spectral band parameters vs. rotation phase}

\begin{figure*}[!ht]
\begin{center}
\includegraphics[height=13cm]{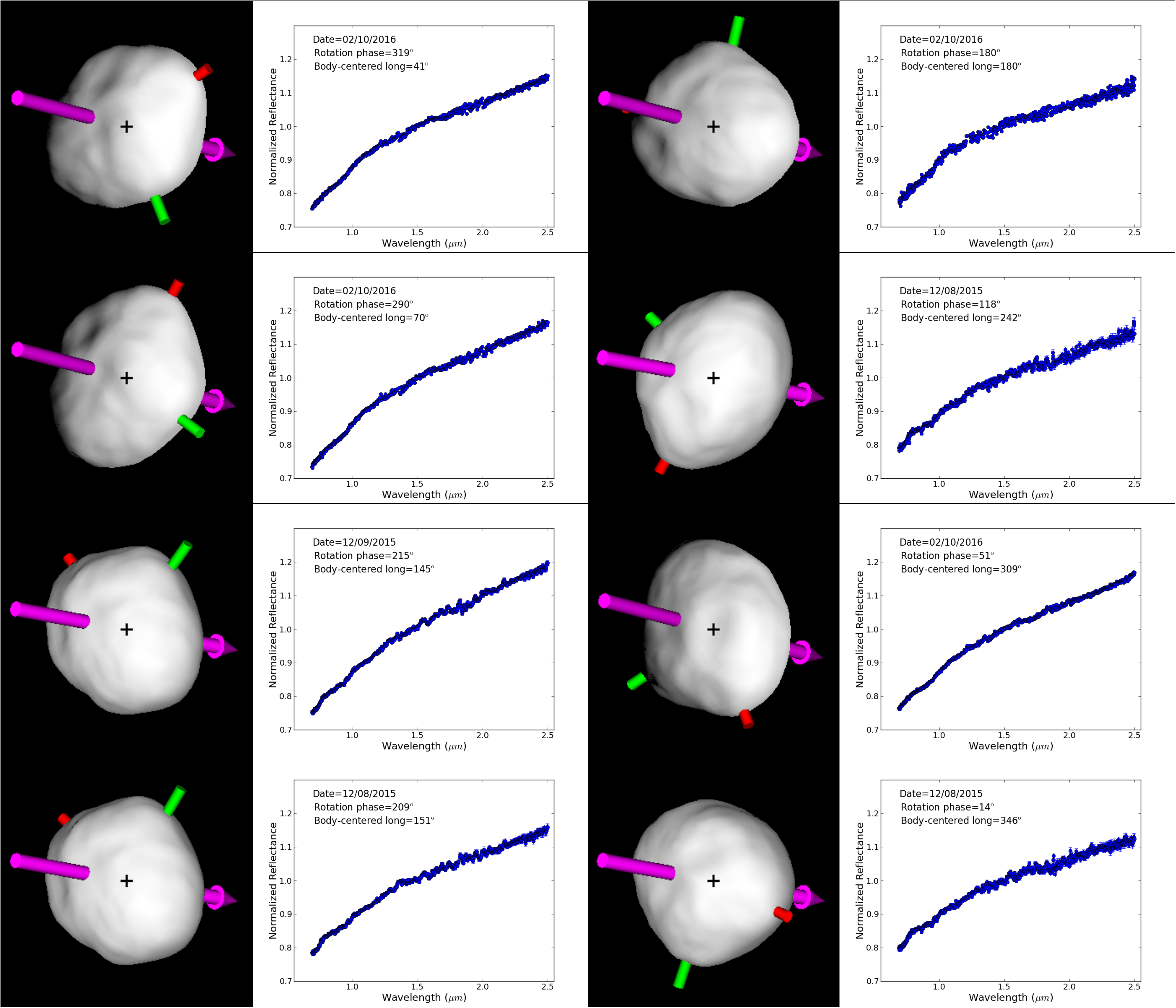}

\caption{\label{f:Allspec2} {\footnotesize NIR spectra and shape models of (16) Psyche corresponding to eight different rotation phases. All spectra are normalized to unity at 
1.5 $\mu m$. The axes shown on the shape models correspond to longitude 0$^\mathrm{o}$ (red) and longitude 90$^\mathrm{o}$ (green).}}

\end{center}
\end{figure*}

NIR spectra of (16) Psyche corresponding to eight different body-centered longitudes are shown in Figure \ref{f:Allspec2}. This figure also includes views of the shape model of Psyche for each rotation phase. \cite{2017Icar..281..388S} 
used 18 Arecibo delay-Doppler radar observations and six new adaptive optics images from Keck and Magellan to generate a three-dimensional shape model of Psyche using the SHAPE software package 
\citep{2007Icar..186..152M}. The shape model incorporates the spin-state and ephemeris so that for any time in the past or future, a plane-of-sky view can be generated to correlate with spectra 
or other telescopically derived data. For the shape model, we adopted a rotation period $P=4.195948\pm0.000001$ h and a spin pole of ($\lambda,\beta$) $(34^\mathrm{o}, -7^\mathrm{o})\pm5^\mathrm{o}$. For each of our 
IRTF runs, we generated plane-of-sky views using the mid-point time of our observations.

\begin{table}[!ht]
\begin{center}\footnotesize
\caption{\label{t:Table2} {\footnotesize Spectral band parameters and composition. The columns in this table are: Sub-Earth latitude, body-centered longitude, band center (BC), band depth (BD), spectral 
slope, molar content of ferrosilite (Fs), molar content of wollastonite (Wo), and orthopyroxene-metal abundance ratio.}}

\rule{0pt}{2ex}

\hspace*{-1.9cm}
\begin{tabular}{crrrrrrrrrrr}
\tableline\tableline

Lat. ($^\mathrm{o}$)&Long. ($^\mathrm{o}$)&BC ($\mu$m)&BD (\%)&Slope ($\mu m^{-1}$)&Fs (mol\%)&Wo (mol\%)&opx/(opx+metal)   \\

-46.74&346&0.940$\pm$0.010&1.3$\pm$0.1&0.25$\pm$0.01&48.2$\pm$3.3&11.8$\pm$1.1&0.06$\pm$0.01 \\
-46.75&242&0.940$\pm$0.005&1.4$\pm$0.1&0.26$\pm$0.01&48.2$\pm$3.3&11.8$\pm$1.1&0.06$\pm$0.01 \\
-46.76&151&0.935$\pm$0.010&1.0$\pm$0.1&0.27$\pm$0.01&43.1$\pm$3.3&9.8$\pm$1.1&0.05$\pm$0.01 \\
-46.99&145&0.940$\pm$0.003&1.2$\pm$0.1&0.31$\pm$0.01&48.2$\pm$3.3&11.8$\pm$1.1&0.05$\pm$0.01 \\
-52.19&70&0.935$\pm$0.004&1.3$\pm$0.1&0.35$\pm$0.01&43.1$\pm$3.3&9.8$\pm$1.1&0.06$\pm$0.01 \\
-52.19&41&0.919$\pm$0.006&1.0$\pm$0.1&0.34$\pm$0.01&26.7$\pm$3.3&3.5$\pm$1.1&0.05$\pm$0.01 \\
-52.19&309&0.919$\pm$0.003&1.3$\pm$0.1&0.31$\pm$0.01&26.7$\pm$3.3&3.5$\pm$1.1&0.06$\pm$0.01 \\
-52.18&180&0.928$\pm$0.006&1.5$\pm$0.1&0.31$\pm$0.01&35.9$\pm$3.3&7.1$\pm$1.1&0.06$\pm$0.01 \\

\tableline
\tableline
\end{tabular}\hspace*{-1.8cm}
\end{center}
\end{table}

\begin{figure*}[!ht]
\begin{center}
\includegraphics[height=10cm]{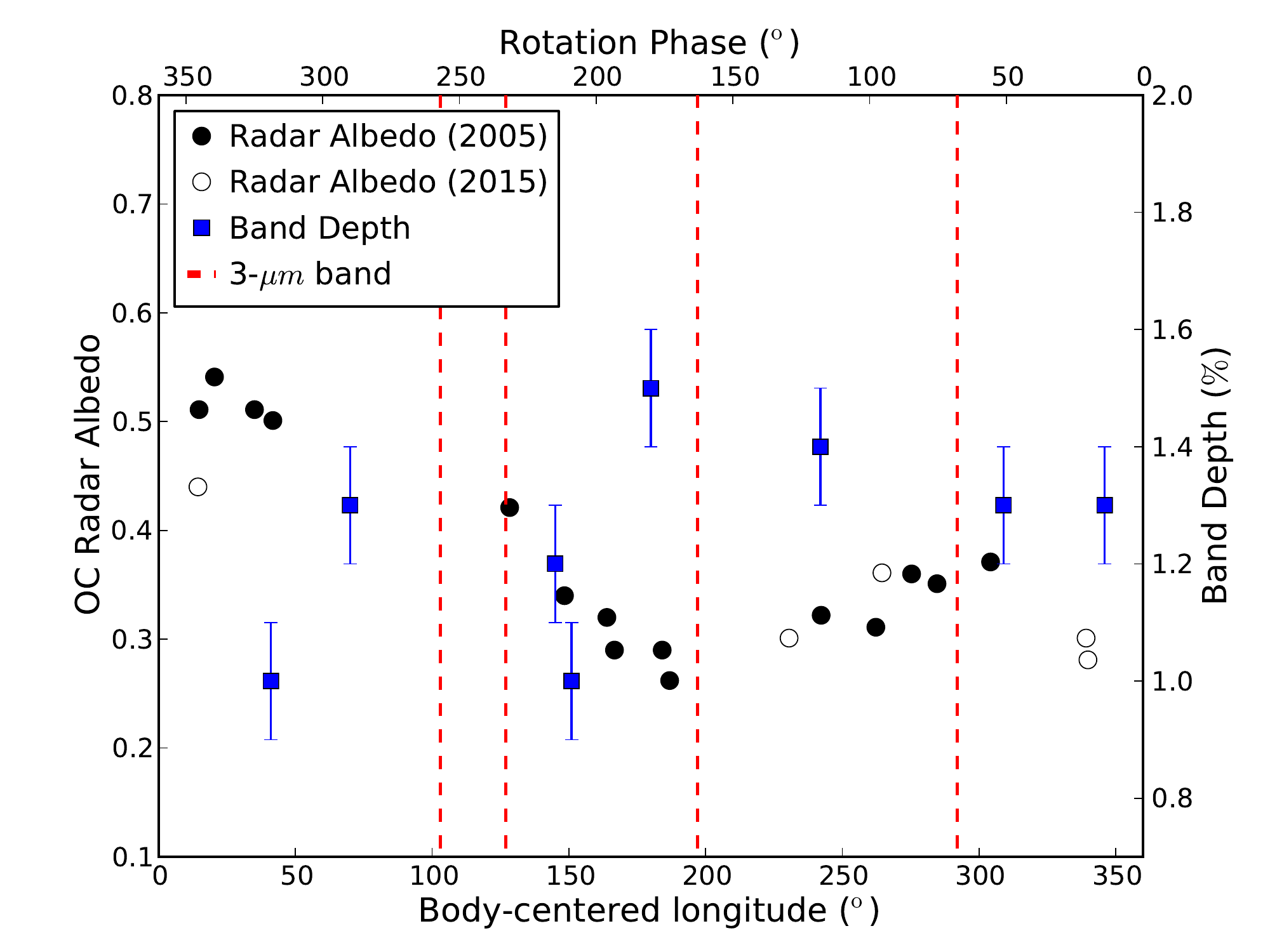}

\caption{\label{f:BD_RPhase2} {\footnotesize Band depth vs. body-centered longitude/rotation phase for (16) Psyche. Radar albedo measurements of Psyche from \cite{2008Icar..195..184S} and 
\cite{2017Icar..281..388S} are also shown. The mean OC radar albedo for Psyche is $\hat{\sigma}_{oc}=0.37\pm0.09$ \citep{2017Icar..281..388S}. The location of the regions where the 3 $\mu m$ band has been 
detected by \cite{2016Nature} is depicted as red dashed lines.}}

\end{center}
\end{figure*}

\begin{figure*}[!ht]
\begin{center}
\includegraphics[height=10cm]{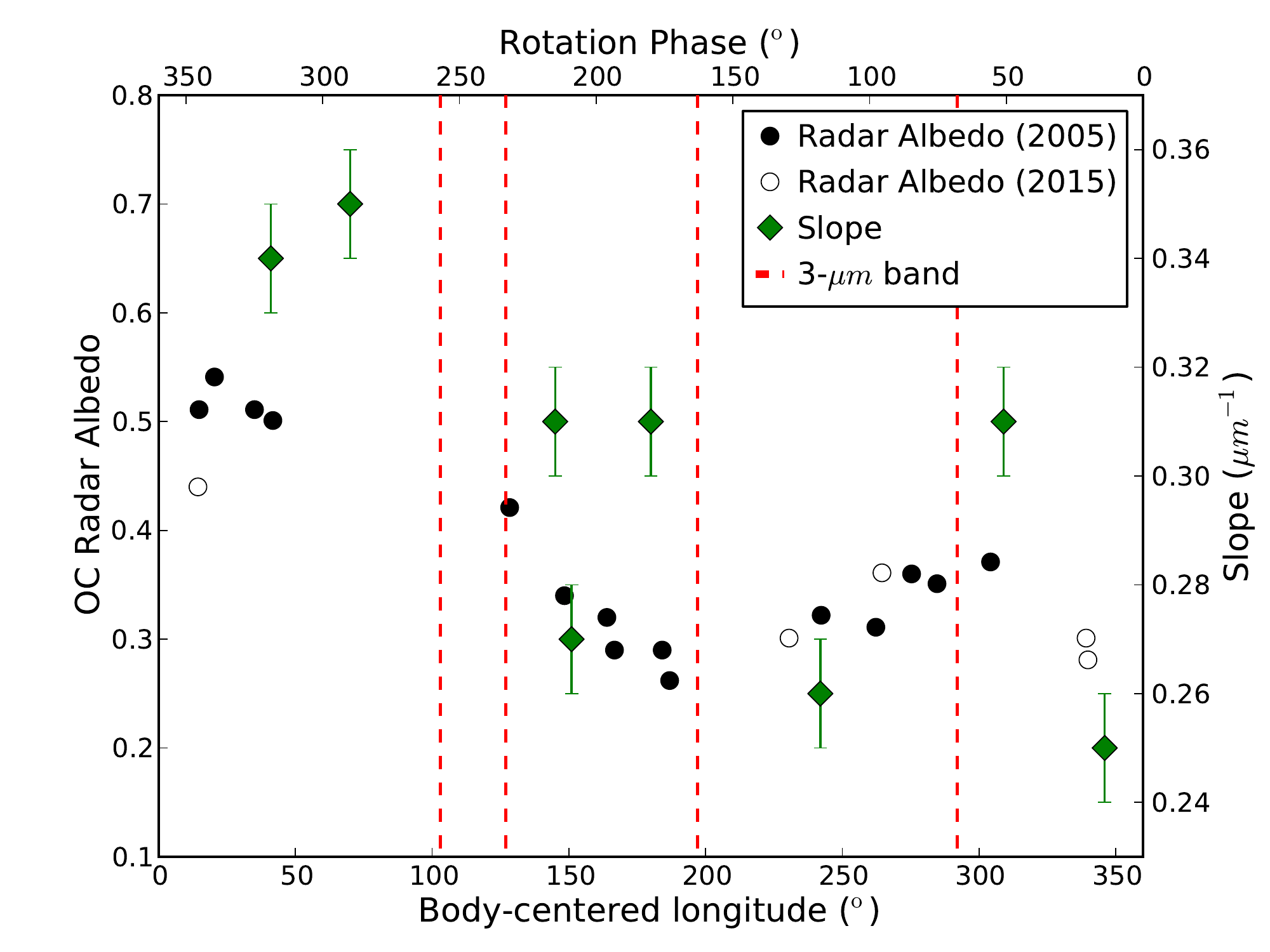}

\caption{\label{f:slope_RPhase2} {\footnotesize Spectral slope vs. body-centered longitude/rotation phase for (16) Psyche. Radar albedo measurements of Psyche from \cite{2008Icar..195..184S} 
and \cite{2017Icar..281..388S} are also shown. The mean OC radar albedo for Psyche is $\hat{\sigma}_{oc}=0.37\pm0.09$ \citep{2017Icar..281..388S}. The location of the regions where the 3 $\mu m$ band has 
been detected by \cite{2016Nature} is depicted as red dashed lines.}}

\end{center}
\end{figure*}

As can be seen in Figure \ref{f:Allspec2}, the spectra look relatively homogeneous across the surface of the asteroid, with all of them showing a weak absorption band at $\sim$ 0.9 $\mu$m. This feature has been detected by previous studies and has been attributed to the presence of low-Fe 
orthopyroxene \citep[e.g.,][]{2004AJ....128.3070C, 2005Icar..175..141H, 2010Icar..210..674O, 2014Icar..238...37N}. For each spectrum, we measured the spectral band parameters, band center, band depth and spectral slope using a Python code and following the procedure described in \citet{1986JGR....9111641C}. The band center was calculated by dividing out the 
linear continuum and fitting a polynomial over the bottom third of the band. The band depth was measured as in \citet{1984JGR....89.6329C} and is given as a percentage depth. The 
spectral slope is defined as the slope of a straight line fitted from 0.76 to 1.2 $\mu$m. Each band parameter was measured ten times by using third- and fourth-order polynomial fits and sampling different ranges of points 
within the corresponding intervals. The reported values correspond to the average of these measurements. Uncertainties are given by the standard deviation of the mean calculated from the ten measurements of 
each parameter. Table \ref{t:Table2} lists the band center, band depth, and spectral slope values measured for each spectrum.

From our measurements we found that the band center varies from $\sim$ 0.92 to 0.94 $\mu$m, with an average value of 0.932$\pm$0.006 $\mu$m, which is consistent with the values 
reported by previous studies \citep[e.g.,][]{2005Icar..175..141H, 2010Icar..210..674O, 2010Icar..210..655F}. Variations of the 
band depth with rotation phase were also observed, with values ranging from 1.0 to 1.5\%, and an average value of 1.3$\pm$0.1\%. Similarly, variations of the spectral slope with the rotation 
phase were found, with minimum and maximum values of 0.25 and 0.35 $\mu m^{-1}$, respectively, and an average slope of 0.30$\pm$0.01 $\mu m^{-1}$. In Figures \ref{f:BD_RPhase2} 
and \ref{f:slope_RPhase2} we plot the band depth and spectral slope versus the rotation phase, along with radar albedo measurements of Psyche from \cite{2008Icar..195..184S} and 
\cite{2017Icar..281..388S}. A possible anti-correlation between band depth and radar albedo (i.e., increasing band depth with decreasing radar albedo) can be seen. 

\cite{2017Icar..281..388S} determined that Psyche has an ellipsoidal shape; 
however, they noticed a deficit of mass at longitudes between 0$^\mathrm{o}$ and 90$^\mathrm{o}$ at the equator. This mass-deficit region, which exhibits the highest 
radar albedo, shows the highest value for the spectral slope and the minimum band depth, while the antipode of this region 
(longitudes $\sim$ 180$^\mathrm{o}$-230$^\mathrm{o}$), where the radar albedo reaches its lowest value, shows the maximum in band depth and less steep spectral 
slopes.

\clearpage

\subsection{Compositional analysis}

\subsubsection{Rotational variations in the pyroxene chemistry}

The variations observed in the band center could be attributed to differences in the pyroxene chemistry across the surface of the asteroid. Previous 
studies \citep[e.g.,][]{2005Icar..175..141H, 2011M&PS...46.1910H} calculated the pyroxene chemistry of only those M-types 
whose spectra exhibit both pyroxene bands (i.e., using the Band I and Band II centers), and since the Band II, which is centered at $\sim$1.9-2 $\mu$m, is not present in the spectrum of 
Psyche, the pyroxene chemistry of this asteroid was never calculated. The pyroxene chemistry is given by the molar content of ferrosilite (Fs) and wollastonite (Wo) and can be 
calculated using \cite{2002aste.conf..183G} or \cite{2007LPI....38.2117B} calibrations. The equations of \cite{2002aste.conf..183G} require both band centers; however, those derived by 
\cite{2007LPI....38.2117B} can be used even if only one pyroxene band is present. Thus, the equations of \cite{2007LPI....38.2117B} were used along with the measured Band I center. 
These equations were derived from the analysis of howardites, eucrites and diogenites (HEDs), and therefore provide only a rough estimation. We determined that the pyroxene chemistry 
for Psyche ranges from Fs$_{26.7\pm3.3}$Wo$_{3.5\pm1.1}$ to Fs$_{48.2\pm3.3}$Wo$_{11.8\pm1.1}$, given an average of Fs$_{40}$En$_{51}$Wo$_{9}$. The calculated values for each rotation phase are included in Table \ref{t:Table2}. This broad range of values must be taken with caution, since it is not clear whether they are the 
result of the scattering of the data or actual surface heterogeneities. Furthermore, the presence of an additional mafic silicate phase (e.g., olivine or high-Ca pyroxene) would shift the band 
center to longer wavelengths, without necessarily introducing a Band II, thereby causing an overestimation of the Fs content. 

\begin{figure*}[!ht]
\begin{center}
\includegraphics[height=10cm]{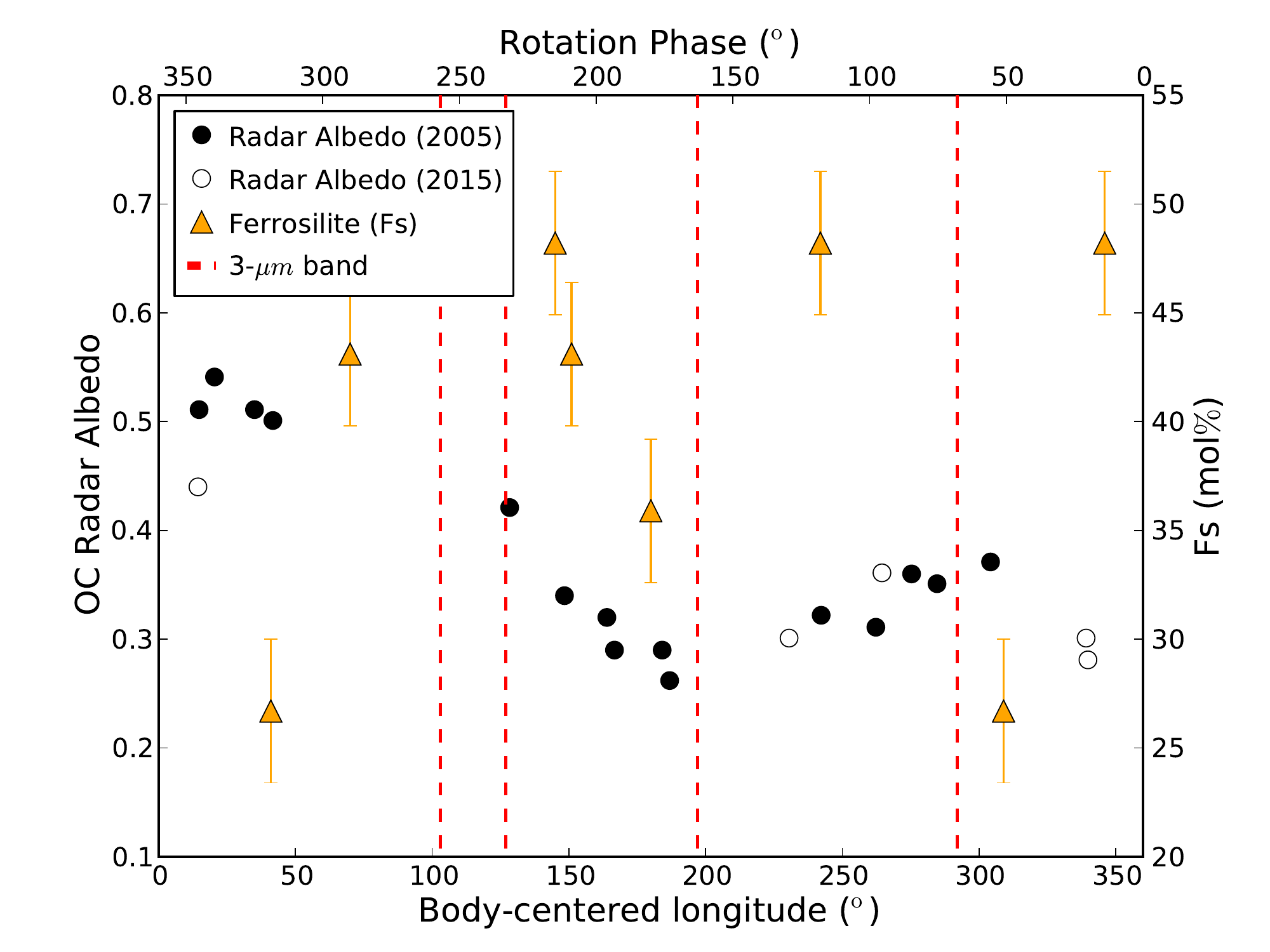}

\caption{\label{f:Fs_RPhase2} {\footnotesize Molar content of ferrosilite (Fs) vs. body-centered longitude/rotation phase for (16) Psyche. Radar albedo measurements of Psyche from 
\cite{2008Icar..195..184S} and \cite{2017Icar..281..388S} are also shown. The mean OC radar albedo for Psyche is $\hat{\sigma}_{oc}=0.37\pm0.09$ \citep{2017Icar..281..388S}. The location of the regions where 
the 3 $\mu m$ band has been detected by \cite{2016Nature} is depicted as red dashed lines.}}

\end{center}
\end{figure*}

\begin{figure*}[!ht]
\begin{center}
\includegraphics[height=11cm]{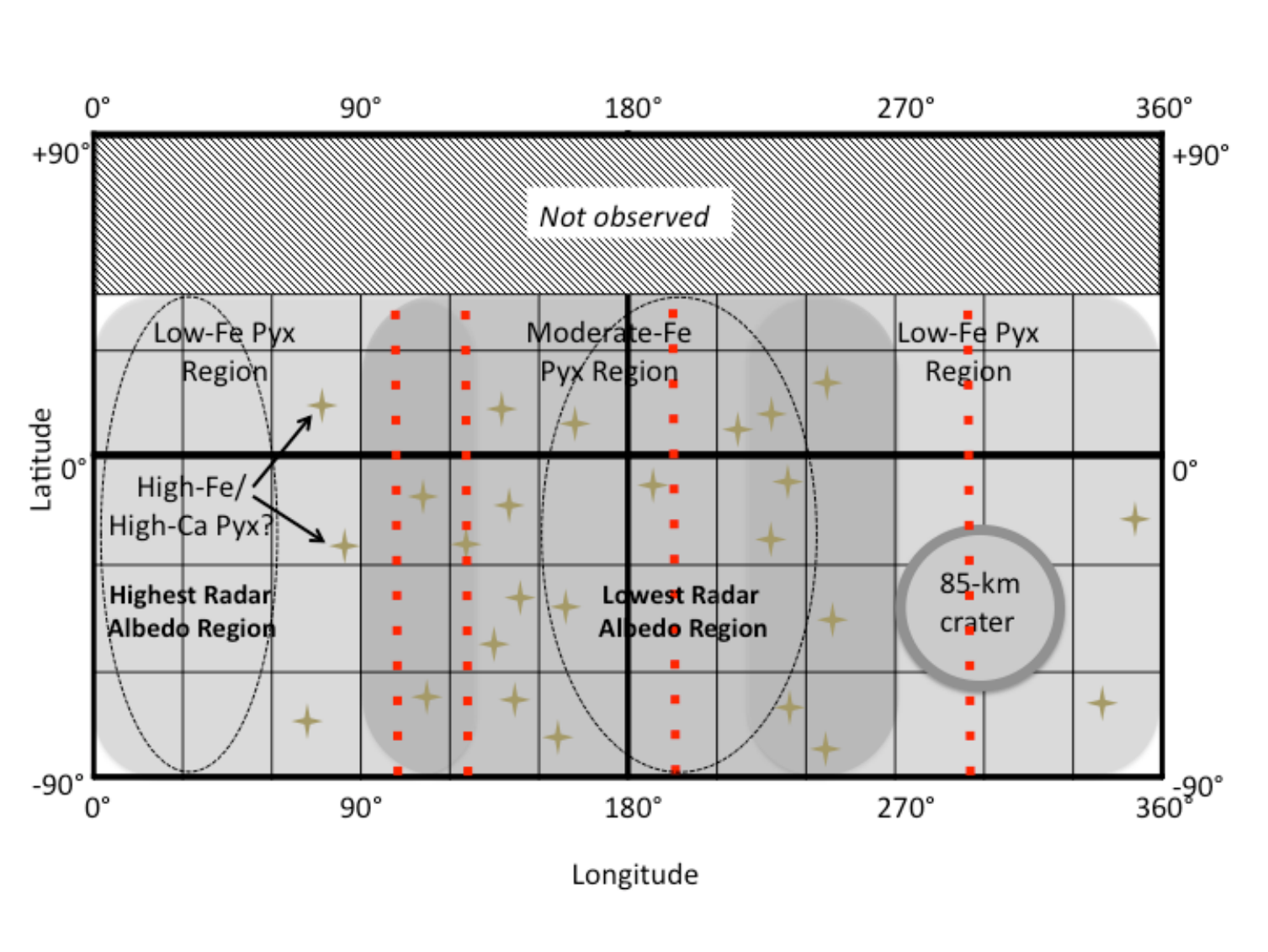}

\caption{\label{f:Map} {\footnotesize Schematic map of (16) Psyche. Low-Fe and moderate-Fe pyroxene (Pyx) regions are depicted in grayscale. The highest and lowest radar albedo regions from 
\cite{2017Icar..281..388S} are indicated as ovals. Star symbols are used to indicate that high-Fe pyroxene or a secondary phase (e.g., high-Ca pyroxene) might be present across the surface of the asteroid. A crater-like 
depression 85$\pm$20 km in diameter located at a body-centered longitude of 300$^\mathrm{o}$ found by \cite{2017Icar..281..388S} is shown. The location of the regions where the 3 $\mu m$ band has been 
detected by \cite{2016Nature} is depicted as red dashed lines.}}

\end{center}
\end{figure*}

Comparing 
these results with those obtained by \citet{2011M&PS...46.1910H}, who studied a large sample of M-type asteroids, we notice that the average surface pyroxene chemistry of Psyche is very similar to that 
calculated for asteroids (338) Budrosa (Fs$_{39}$En$_{53}$Wo$_{8}$) and (497) Iva (Fs$_{41}$En$_{50}$Wo$_{9}$). However, in contrast to those of Psyche, the spectra of these two asteroids 
exhibit both pyroxene bands. According to \citet{2011M&PS...46.1910H}, the position of the band centers of Budrosa and Iva in the Band I vs. Band II plot, "{\it{slightly above the pyroxene trend}}", could be attributed to 
the presence of either an olivine or a high-Ca pyroxene phase. The presence of these minerals could then lead to an overestimation of the Fs abundance. Thus, it is possible that there are no regions on Psyche that are 
more ferrous than others, but contain residues of other minerals, like high-Ca pyroxene. Considering this, the lowest calculated values are likely more representative of the pyroxene chemistry of Psyche. 
Therefore, if we recalculate the average considering only the lowest values of Fs and Wo, corresponding to longitudes of 41$^\mathrm{o}$, 180$^\mathrm{o}$ and 309$^\mathrm{o}$, we obtain Fs$_{30}$En$_{65}$Wo$_{5}$. This value is within the range found for pyroxene megacrysts and the mafic melt matrix of mesosiderite meteorites \citep[e.g.,][]{1997LPI....28..125B}. Figure \ref{f:Fs_RPhase2} shows 
the molar content of Fs as a function of body-centered longitude for Psyche. The lowest value of Fs is found for body-centered 
longitudes of 41$^\mathrm{o}$ and 309$^\mathrm{o}$, with the former corresponding to a possible excavated region, perhaps similar to Vesta's Rheasilvia basin \citep{2017Icar..281..388S}, and the latter corresponding 
to the location of a crater-like depression 85$\pm$20 km in diameter found by \cite{2017Icar..281..388S}. We notice that out of eight, five rotation phases exhibit high Fs content. If these high values are 
indeed the result of an overestimation caused by a secondary or accessory mineral, then this mineral is likely present across the surface of the asteroid, with only a few zones (possible excavated regions) with a minor 
content of it. A schematic map of Psyche is shown in Figure \ref{f:Map}. All the spectral observations were obtained at a sub-Earth latitude of $\sim$ 50$^\mathrm{o}$S. Low-Fe and moderate-Fe pyroxene regions are depicted 
in grayscale.

\subsubsection{Rotational variations in metal abundance}

Variations in band depth and spectral slope could be the result of differences in mineral abundance. In order to explore this possibility, we calculate the fraction of pyroxene and metal 
present as a function of rotation phase. We choose these two minerals because they are the ones that better fit the spectral characteristics of Psyche. To this end, we have developed a 
new laboratory spectral calibration using data from \cite{2009LPI....40.1332C}, who measured the reflectance spectra of intimate mixtures composed of orthopyroxene (Fs$_{12.8}$) and metal 
from the Odessa octahedrite iron meteorite (grain size $<$ 45 $\mu$m in both cases). The reflectance spectra of the samples correspond to intimate mixtures where the orthopyroxene 
abundance was increased in 10 wt.\% intervals. 

\begin{figure*}[!ht]
\begin{center}
\includegraphics[height=9cm]{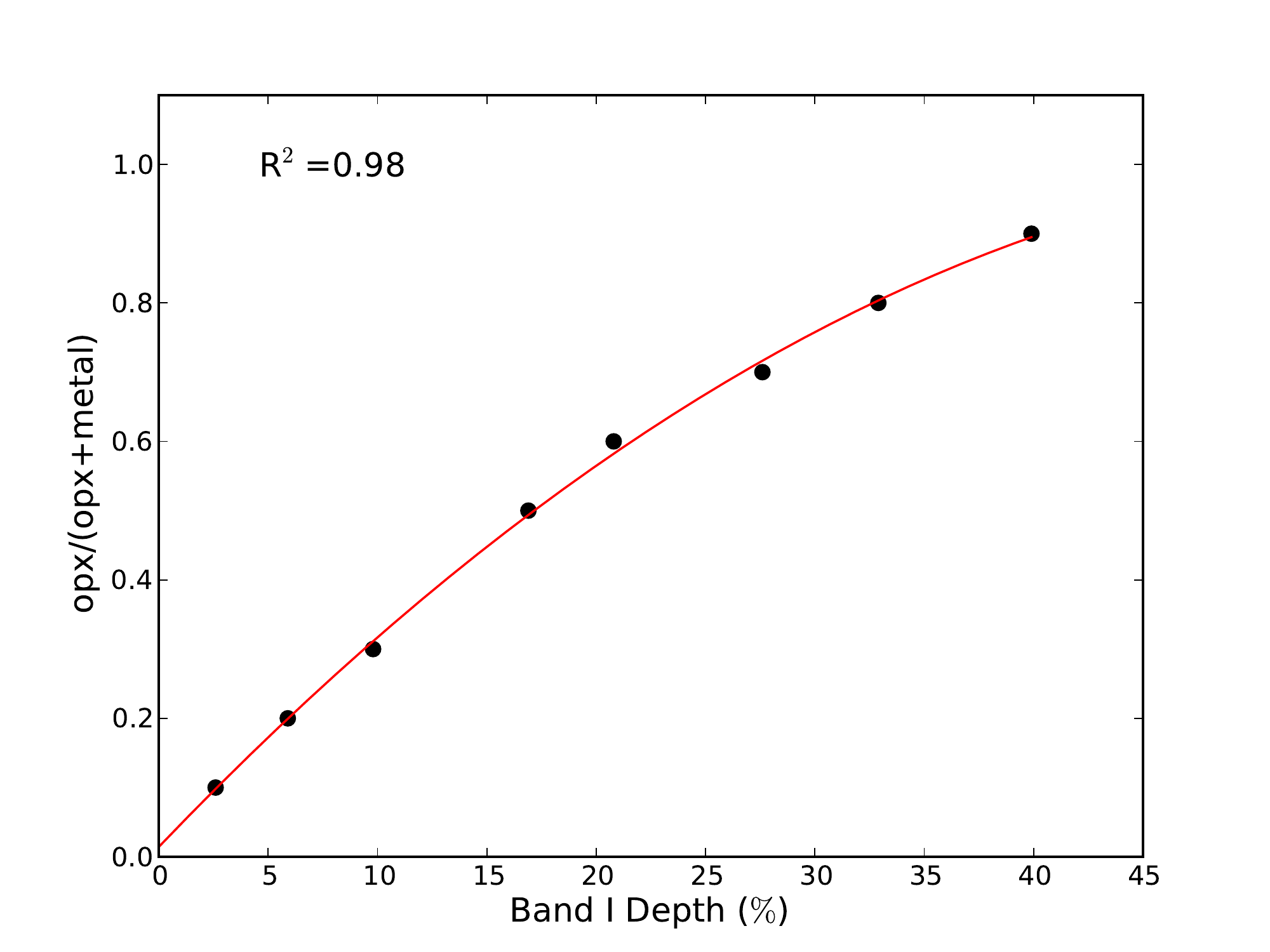}

\caption{\label{f:polyfit} {\footnotesize opx/(opx + metal) ratios vs. band depth for intimate mixtures composed of orthopyroxene and metal 
from the Odessa octahedrite iron meteorite. The red line represents a second order polynomial fitted to the data. The coefficient of determination (R$^{2}$) is given. The root mean square error 
between the opx/(opx+metal) ratio determined using Eq. (1) and the real values is 0.01.}}

\end{center}
\end{figure*}

For each spectrum, we measured the band I depth in the same way as we did it for the asteroid spectra. Figure \ref{f:polyfit} shows the orthopyroxene-metal abundance 
ratio (opx/(opx + metal)) versus band depth for the intimate mixtures. As noted by \cite{2009LPI....40.1332C}, an increase in the metal content will cause the band depth to decrease 
and the spectral slope to increase. We found that the correlation between the opx/(opx+metal) and band depth (BD) can be described by 
the following second-order polynomial fit:

\begin{equation}
opx/(opx+metal)=-0.000274\times BD^{2}+0.033\times BD+0.014
\end{equation}

where the coefficient of determination (R$^{2}$) is 0.98 and the root mean square error between the opx/(opx+metal) ratio determined using Eq. (1) and the real values is 0.01. Using this 
equation along with the band depths measured from each spectrum of Psyche, we determined the pyroxene abundance. These values are included in Table \ref{t:Table2}. We found that the 
maximum variation of pyroxene content with the rotation phase is 1 \%, with an average opx/(opx+metal) ratio of 6$\pm$1\%. The lowest opx/(opx+metal) ratio corresponds to longitudes of  
41$^\mathrm{o}$, 145$^\mathrm{o}$, and 151$^\mathrm{o}$. These results indicate that a small difference in metal content could explain the variations of the band depth and spectral slope 
with the rotation phase. However, since this difference is on the order of the uncertainty associated with the method we are using, we cannot be completely certain.

\begin{figure*}[!ht]
\begin{center}

\includegraphics[height=9cm]{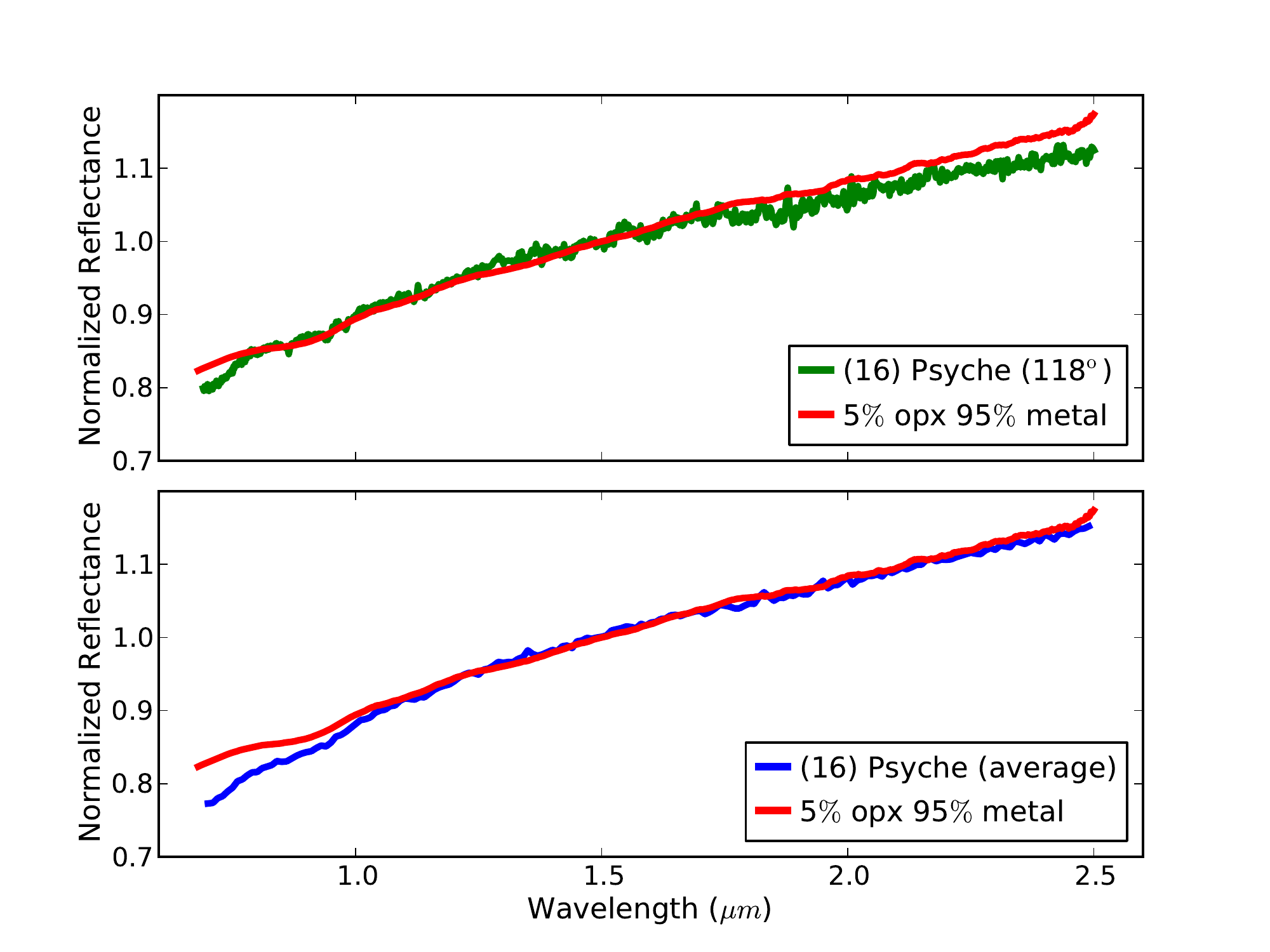}

\caption{\label{f:psyche_opxmetal2} {\footnotesize Comparison of the spectrum of an intimate mixture of 5 wt.$\%$ orthopyroxene and 95 wt.$\%$ metal from 
\cite{2009LPI....40.1332C} to the spectrum of Psyche corresponding to rotation phase 118$^\mathrm{o}$ (top) and to the average spectrum of Psyche (bottom). All spectra are 
normalized to unity at 1.5 $\mu m$.}}

\end{center}
\end{figure*}

For comparison, in Figure \ref{f:psyche_opxmetal2} we show the spectrum of an intimate mixture of 5 wt.$\%$ orthopyroxene and 95 wt.$\%$ metal from 
\cite{2009LPI....40.1332C}. The overall shape of the spectrum of this intimate mixture is very similar to that of Psyche spectra, although some individual spectra fit better than 
others. In this example, we can see how the spectrum corresponding to rotation phase 118$^\mathrm{o}$ (longitude 242$^\mathrm{o}$) is more similar to that of the 
intimate mixture than the average spectrum of Psyche, whose spectral slope is slightly different. Since we have normalized the spectra to unity at 1.5 $\mu 
m$ the difference in spectral slope is seen as a drop in reflectance shortward of 1.2 $\mu m$. 

\clearpage

\subsection{Effects of phase angle, grain size and space weathering}

Variations in band depth and spectral slope could be attributed to factors other than mineral abundance, including differences in phase angle, grain size, and space weathering. As can 
be seen in Table \ref{t:Table1}, observations of Psyche were obtained at two very different phase angles, 1.8 and 19.5$^\mathrm{o}$. An increase in phase angle can produce changes in the 
strength of the absorption band and increase the spectral slope \citep{2012Icar..220...36S} and could therefore explain the variations of these two parameters between different observing 
runs. However, this cannot explain the variations seen within each observing run. This opens the possibility for variations of grain size and space weathering with the rotation phase. 

An increase in grain size will typically produce an increase in band depth, albedo variations, and a decrease in spectral slope \cite[e.g.,][]{2009LPI....40.1332C, 2010M&PS...45..304C, 2015Icar..252...39C}. This alternative 
explanation would be compatible with the albedo variations across the surface of Psyche reported by previous studies \cite[e.g.,][]{2002Icar..159..369K}. Furthermore, a change in the surface texture is considered among the 
possible causes for the variations in radar albedo detected on Psyche \citep{2008Icar..195..184S}. 

In the case of lunar-style space weathering, the effects on the spectra can be seen as an increase in spectral slope and a suppression of the absorption bands 
\cite[e.g.,][]{2000M&PS...35.1101P, 2001JGR...10610039H, 2006Icar..184..327B, 2010Icar..209..564G}. Thus, rotation phases whose spectra exhibit less steep slopes and deeper absorption bands could represent 
regions on the surface where fresh material has been exposed. In addition, there is new evidence that suggests the presence of H$_{2}$O/OH-bearing minerals on the surface of Psyche. In a recent study, \cite{2016Nature} 
detected a 3 $\mu m$ band in the mid-IR spectra (1.9-4.1 $\mu m$) of Psyche. The locations where this band was detected are depicted in Figures \ref{f:BD_RPhase2} to \ref{f:Map} (red dashed lines). This 3 $\mu m
$ band could be related to the presence of dark exogenic material (e.g., carbonaceous chondrites), like the one found on the surface of Vesta \cite[e.g.,][]{2012Icar..221..544R}. If this is the case, then the presence of the dark 
material could have also contributed to the variations seen in the spectral slope and band depth.

\section{Discussion}

\begin{figure*}[!ht]
\begin{center}

\includegraphics[height=10cm]{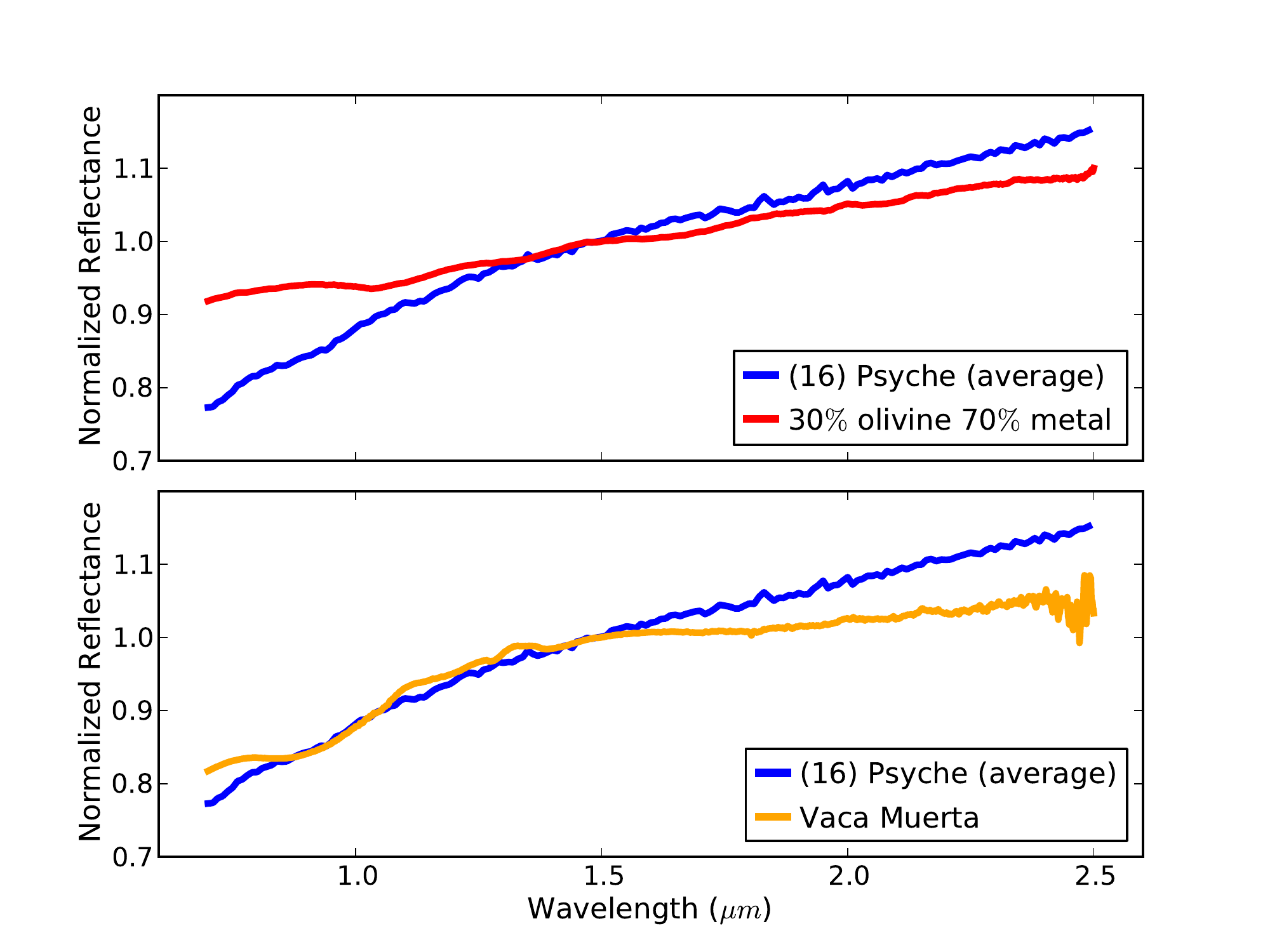}

\caption{\label{f:Vmuerta2} {\footnotesize Top: NIR spectrum of an intimate mixture of 30 wt.$\%$ olivine and 70 wt.$\%$ metal from \cite{2015Icar..252...39C} 
compared to the average spectrum of Psyche. Bottom: NIR spectrum of the mesosiderite Vaca Muerta obtained at the University of Winnipeg HOSERLab compared to the average 
spectrum of Psyche. Differences between the two spectra at wavelengths $>$ 1.5 $\mu m$ could be attributed to different grain sizes and metal abundance. Space weathering and/or 
differences in metal grain size likely account for the spectral mismatch below $\sim$0.85 $\mu m$. All spectra are normalized to unity at 1.5 $\mu m$.}}

\end{center}
\end{figure*}

The conditions that led to the formation of Psyche remain a mystery. As stated earlier, one of the most accepted theories is that this object is the exposed metallic core of 
an $\sim$ 500 km diameter differentiated body. The structure of a differentiated body, which has been molten and density-segregated, consists of an FeNi-FeS core, an 
olivine-rich mantle, and a pyroxene-feldspar (basaltic) crust \cite[e.g.,][]{1989aste.conf..921B, 1993Icar..106..573G}. Our analysis showed a relatively homogeneous metal content 
across the surface of Psyche, with only a small fraction ($\sim$ 6$\%$) of silicates. The silicate component, which is dominated by orthopyroxene, seems to be intimately mixed with the metal, as indicated by 
the presence in all of the spectra of a weak pyroxene absorption band. The possible presence of a secondary phase, possibly high-Ca pyroxene, across the surface suggests that this mineral was originally derived 
from the parent body. An interesting aspect about Psyche is the apparent lack or low content of olivine on the surface, as revealed by our observations and previous studies. Numerical simulations have 
shown that hit-and-run collisions can strip the silicate crust and mantle from differentiated bodies \cite[e.g.,][]{2006Natur.439..155A, 2014NatGe...7..564A}. However, it is unlikely that this event 
would produce a pure metallic core, without any residual olivine mantle. Subsequent collisions and micrometeorite impacts would shatter the residual silicate 
fragments, producing a regolith layer mainly composed of metal and a small fraction of olivine. Figure \ref{f:Vmuerta2} (top panel) shows an example of the NIR spectrum of an 
intimate mixture of 30 wt.$\%$ olivine and 70 wt.$\%$ metal from \cite{2015Icar..252...39C}. This spectrum exhibits a weak feature centered at $\sim$1.05 $\mu$m, very different 
from Psyche spectra, whose band center varies from $\sim$ 0.92 to 0.94 $\mu$m. This suggests that Psyche's parent body did not have the typical structure expected for a 
differentiated body with an oxidized precursor or that the sequence of events that led to Psyche's current state was more complex than expected. 

For planetesimals with different chondritic compositions that underwent melting and differentiation, the first crystallizing mineral in almost all cases is olivine, and therefore, this is 
the most likely silicate remnant to be present on the surface of Psyche \citep{2013LPI....44.1351E}. However, if the planetesimal had a silica-rich bulk composition (more reduced) and the 
pressure of the interior was sufficiently low, then the first crystallizing mineral would have been pyroxene \citep{2013LPI....44.1351E}. If this was the case for Psyche's parent body, then this 
could explain the apparent lack of olivine on the surface of this asteroid. 

Another possible explanation could involve a formation 
scenario similar to the one proposed for 
the mesosiderites' parent body. Mesosiderites are stony-iron meteorites composed of approximately equal amounts of Fe,Ni-metal, and silicates plus troilite 
\citep{2006mess.book...19W}. The silicate component is similar in composition to HED meteorites, containing olivine, pyroxenes, and 
plagioclase \citep{1998LPI....29.1220M}. One of the interesting features of these meteorites is that crustal and core materials are abundant, while olivine from the mantle is rare. 
Figure \ref{f:Vmuerta2} (bottom panel) shows the NIR spectrum of a slab of the mesosiderite Vaca Muerta. As can be seen in this figure, the overall shape of the spectrum is similar 
to that of Psyche, with a pyroxene absorption band centered at $\sim$0.91 $\mu$m. The differences in overall slope beyond $\sim$1.5 $\mu$m can be attributed to differences in metal 
grain size: Psyche likely has a powdered regolith, which would lead to a more red-sloped spectrum as compared to our mesosiderite spectra, which were measured as points on a 
roughened slab. For metal+mafic silicates, slab spectra are less red-sloped than powders. For mesosiderite slab spectra, increasing metal content (as determined visually where spot 
spectra were acquired) results in more red-sloped spectra and shallower pyroxene absorption band depths, as expected. Space weathering would have the greatest effect at the lowest 
wavelengths, causing a reddening of the spectra slope. This is evident in the Psyche spectra versus the laboratory spectra of the mineral mixtures and mesosiderite slab spectra.

Numerical experiments performed by \citet{2001M&PS...36..869S} showed that mesosiderites could have been formed by the disruption and reassembly of a Vesta-like asteroid. In this scenario, the 
target, a differentiated asteroid with a molten core, is disrupted by a smaller projectile. According to \citet{2001M&PS...36..869S}, after the 
disruption event most of the molten metal would have crystallized around the small, cool crustal fragments, as they were more efficient at trapping molten metal than the hotter 
and larger fragments from the mantle. As a result, olivine-rich mantle material would have been preferentially excluded from the metal-rich regions of the reaccreted body. The 
lower survival rate of the olivine fragments could have also contributed to reducing the olivine content in the reaccreted body, which would consist of a mixture of metal, pyroxenite, 
basalt, and dunite fragments. After the new reaccreted body was formed, subsequent collisions must have taken place, but this time with the reaccreted 
body acting as the impactor. The reason is that in a hit-and-run collision the impactor is more easily disrupted than the target it strikes, losing most of the outer layers in the process 
with no subsequent accretion \citep{2006Natur.439..155A, 2014NatGe...7..564A}. These hit-and-run collisions would result in an exposed metallic core, probably retaining only a 
small fraction of silicates composed of intimately mixed materials from all depths. We speculate that this sequence of events could also explain the spectral characteristics seen on 
Psyche.

\section{Summary}

We have obtained rotationally resolved NIR spectra of the asteroid (16) Psyche. Band parameters were measured from each spectrum corresponding to different rotation phases. We found 
variations of the band center (0.92-0.94 $\mu$m), band depth (1.0-1.5\%) and spectral slope (0.25-0.35 $\mu m^{-1}$) with the rotation phase. A possible anti-correlation between the band depth and radar albedo is observed. The mass-deficit region, characterized as having the highest radar albedo, also shows the highest value for the spectral slope and the minimum band depth, while 
the antipode of this region, where the radar albedo reaches its lowest value, shows the maximum in band depth and less steep spectral slopes. 

Band center values were used to calculate the pyroxene chemistry of Psyche, which ranges from Fs$_{26.7\pm3.3}$Wo$_{3.5\pm1.1}$ to Fs$_{48.2\pm3.3}$Wo$_{11.8\pm1.1}$. The lowest value of Fs is found for 
body-centered longitudes of 41$^\mathrm{o}$ and 309$^\mathrm{o}$, corresponding to possible excavated regions, including a crater-like depression 85$\pm$20 km in diameter located at a longitude of 
300$^\mathrm{o}$. The highest values of Fs could be the result of an overestimation caused by the presence of a secondary or accessory mineral (e.g., olivine or high-Ca pyroxene). We determined that an average 
value of Fs$_{30}$En$_{65}$Wo$_{5}$ is likely more representative of the pyroxene chemistry of Psyche. 

Using a new laboratory spectral calibration we estimated an average pyroxene abundance of 6$\pm$1\%. The spectral characteristics of Psyche on a global scale suggest that the metal on the 
surface and the silicate component are well mixed, as indicated by the presence in all of the spectra of a weak pyroxene absorption band in the 0.9 $\mu$m region. Furthermore, the distribution of a secondary 
phase, possibly high-Ca pyroxene, across the surface of Psyche suggests that this is an indigenous mineral derived from the parent body.

The apparent lack of olivine on the surface of Psyche could indicate that pyroxene and not olivine was the first crystallizing mineral in the mantle of the parent body. Another possible explanation could involve the 
disruption and reaccretion of a differentiated asteroid with a molten core, followed by subsequent hit-and-run collisions of the reaccreted body. In this scenario, the 
olivine fraction is reduced during the reaccretion process, and the hit-and-run collisions leave behind an exposed metallic core covered by a regolith layer of metal and pyroxene.

The effects of the phase angle, grain size, and space weathering on the spectral data were examined. We observed that these factors could be responsible to some extent for the 
variations seen in the band parameters. However, since they are difficult to disentangle, it is not possible to know how significant their contributions are.

\acknowledgments

\

\begin{center}

{\bf{Acknowledgements}}

\end{center}
\
This research work was supported by NASA Planetary Mission Data Analysis Program Grant NNX13AP27G, NASA NEOO Program Grant NNX12AG12G, and NASA Planetary Geology 
and Geophysics Program Grant NNX11AN84G. We thank the IRTF TAC for awarding time to this project, and we thank the IRTF TOs and MKSS staff for their support. The IRTF is operated by the 
University of Hawaii under Cooperative Agreement no. NCC 5-538 with the National Aeronautics and Space Administration, Office of Space Science, Planetary Astronomy Program. Part of 
this work was done at the Arecibo Observatory, which is operated by SRI International under a cooperative agreement with the National Science Foundation (AST-1100968) and in alliance 
with Ana G. Mendez-Universidad Metropolitana and the Universities Space Research Association. The Arecibo Planetary Radar Program is supported by the National Aeronautics and 
Space Administration under Grant Nos. NNX12AF24G and NNX13AQ46G issued through the Near Earth Object Observations program. EAC thanks the Canada Foundation for Innovation, the Manitoba 
Research Innovations Fund, the Canadian Space Agency, the Natural Sciences and Engineering Research Council of Canada, and the University of Winnipeg for supporting the establishment and ongoing 
operation of the University of WinnipegÕs Planetary Spectrophotometer Facility. The authors would like to thank the anonymous reviewer for their comments, which helped to improve the manuscript. We also thank Kim Tait and 
Ian Nicklin from the Royal Ontario Museum for providing a suite of mesosiderite samples for this study.

\clearpage

\end{document}